\documentclass{article}

\usepackage[preprint]{neurips_2026}

\usepackage[utf8]{inputenc} 
\usepackage[T1]{fontenc}    
\usepackage{hyperref}       
\usepackage{url}            
\usepackage{booktabs}       
\usepackage{amsmath}        
\usepackage{amsfonts}       
\usepackage{nicefrac}       
\usepackage{microtype}      
\usepackage{xcolor}         
\usepackage{amsthm}        
\usepackage{amssymb}        

\theoremstyle{plain}
\newtheorem{theorem}{Theorem}[section]

\theoremstyle{definition}
\newtheorem{definition}[theorem]{Definition}

\theoremstyle{remark}

\usepackage[textsize=tiny]{todonotes}
\usepackage{xspace}
\def\eg{\emph{e.g.}\xspace} 
\def\ie{\emph{i.e.}\xspace}

\def\etal{\emph{et al.}\xspace}

\usepackage{color}

\title{AI Agents Alone Are Not (Yet) Sufficient\\
for Social Simulation}

\author{%
  \textbf{Yiming Li},
  \textbf{Dacheng Tao} \\
  College of Computing and Data Science, Nanyang Technological University, Singapore\\
  \texttt{liyiming.tech@gmail.com}
}

\begin{document}

\maketitle

\begin{abstract}
Recent advances in large language models (LLMs) have spurred growing interest in using LLM-integrated agents for social simulation, often under the implicit assumption that realistic population dynamics will emerge once role-specified agents are placed in a networked multi-agent setting. This position paper argues that LLM-based agents alone are not (yet) sufficient for social simulation. We attribute this over-optimism to a systematic mismatch between what current agent pipelines are typically optimized and validated to produce and what simulation-as-science requires. Concretely, role-playing plausibility does not imply faithful human behavioral validity; collective outcomes are frequently mediated by agent–environment co-dynamics rather than agent–agent messaging alone; and results can be dominated by interaction protocols, scheduling, and initial information priors. To make these underlying mechanisms explicit and auditable, we propose a unified formulation of AI agent-based social simulation as an environment-involved Markov game with explicit exposure and scheduling mechanisms, from which we derive concrete actions for design, evaluation, and interpretation. 
\end{abstract}

\vspace{-1.5em}
\section{Introduction}
\vspace{-0.5em}
\label{sec:intro}

LLM-integrated agents are increasingly used to construct interactive multi-agent systems mediated by natural language. A rapidly growing line of work proposes to use such agent societies as social simulators, where agents serve as proxies for human individuals and aggregate outcomes are used to study diffusion, polarization, market behavior, and policy interventions \cite{gao2024large,mou2024individual, piao2025agentsociety}. This paradigm is appealing because it appears to enable scalable and repeatable experimentation in settings where human studies are costly, slow, or ethically constrained \cite{ park2023generative,argyle2023out,anthis2025position}.

This optimism has been amplified by a wave of recent demonstrations in which LLM-integrated agents appear strikingly human-like and exhibit emergent society-like patterns at scale. Generative-agent towns produce believable daily routines, gossip, and relationship formation \citep{park2023generative}; large-scale platforms simulate millions of social-media users with longitudinal behavioral traces \citep{piao2025agentsociety}; and most strikingly, agent-only social networks such as \texttt{Moltbook}\footnote{\url{https://www.moltbook.com/}} have reportedly produced spontaneous governance structures, token economies, tribal identities, and even an emergent religion among tens of thousands of LLM agents \citep{zhang2026agents}. Such demonstrations have, understandably, encouraged the inference that role-playing fidelity is converging towards human fidelity, and that placing such agents in a network is itself sufficient to instantiate a society. However, much current practice risks over-interpreting these outputs. Social simulations are often instantiated as role-playing pipelines: personas are assigned, agents exchange messages, and transcript plausibility is treated as evidence of realism \cite{park2023generative,gao2024large,piao2025agentsociety}. This implicitly assumes that human-like local interactions are sufficient to justify collective validity. The assumption becomes particularly fragile when simulations are used to support comparative or counterfactual claims, where conclusions depend not only on interaction quality but also on how information, incentives, and constraints are generated and propagated.

\textbf{Our position is that AI agents alone are not (yet) sufficient for social simulation.} The core issue is epistemic: LLM-agent pipelines are typically optimized and validated for role consistency and conversational plausibility, whereas simulation-as-science requires faithful decision processes and well-specified generative mechanisms. In practice, we find that collective outcomes are frequently shaped by agent–environment co-dynamics, such as exposure mechanisms, institutional constraints, scheduling effects, and information asymmetries. These factors are often under-specified or treated as implementation details, yet they can dominate downstream trajectories and drive apparent effects \cite{park2023generative,ju2024sense,yang2024oasis}. Building on this diagnosis, we identify recurring mismatches between proxy objectives and epistemic goals in current LLM-based social simulation, highlight their concrete gaps, and propose a unified formulation of environment-involved agent-based social simulation. We conclude by calling for a set of practical actions aimed at improving reliability, interpretability, and epistemic clarity before such simulators are treated as scientific or policy-facing evidence.

\vspace{-0.6em}
\section{AI Agents for Social Simulation}
\vspace{-0.4em}

\subsection{Scope}
\vspace{-0.3em}

Social simulation refers to the use of computational models to study how individual cognition, decision-making, and interaction give rise to collective social phenomena \cite{gilbert2005simulation}. In recent years, this paradigm has attracted renewed interest through the use of large language model (LLM)-integrated AI agents, which promise scalable, flexible proxies for human individuals in settings where real-world experimentation is costly, slow, or ethically constrained \cite{gao2024large,mou2024individual,piao2025agentsociety}. By enabling open-ended communication and role-conditioned behavior, LLM-based agents are often viewed as a natural substrate for simulating complex social dynamics in domains such as policy analysis, opinion dynamics, and institutional design \cite{park2023generative,argyle2023out,huang2026dualmind}.


In this paper, we focus on (LLM-integrated) AI agents used explicitly as proxies for human individuals in social simulation, where simulation outcomes are interpreted at the collective level\footnote{More details are in Appendix \ref{sec:appendix}.}. Our analysis centers on the methodological assumptions and epistemic implications of such simulations, rather than on task performance or predictive accuracy. We do not aim to provide a comprehensive survey of social simulation methods, nor do we examine non–agent-based approaches or agent applications outside social simulation. This scoped focus allows us to critically examine prevailing assumptions and overlooked issues of AI agents specifically within the context of social simulation.

\vspace{-0.5em}
\subsection{Current Approaches}
\label{sec:current}
\vspace{-0.3em}

To ground the discussion, we briefly outline how recent work employs LLM-integrated agents for social simulation. Building on existing taxonomies of LLM-based agent simulation \cite{gao2024large,mou2024individual}, we organize current practice along two axes. Most existing pipelines share an implicit premise: sufficiently human-like individual agents, once placed in a networked setting, will yield realistic collective dynamics. In practice, this premise is instantiated along two complementary axes. Some work prioritises \emph{individual-level} simulation, configuring agents as human proxies via personas, memory, and interaction protocols to ensure local plausibility. Other work emphasises \emph{collective-level} simulation, focusing on how agents are composed, connected, and interact to generate and interpret system-level outcomes. This section adopts this decomposition to organize current practice.

\vspace{-0.5em}
\subsubsection{Individual-level Simulation} 
\vspace{-0.2em}

At the individual level, LLM-integrated agents are designed as stable and plausible human proxies. Based on their implementation techniques, these methods can be grouped into two categories. 

\textbf{Prompt-based methods} encode personas in the prompt to shape agent behavior, without updating model parameters. Some curate personas via human-in-the-loop profiling, drawing on well-known characters \cite{yuan2024evaluating} or structured attributes (\eg, gender) \cite{wang2025user}. Others synthesize personas by prompting LLMs with individual details \cite{tu2023characterchat}.


\textbf{Training-based methods} cover pre-training, fine-tuning, and reinforcement learning (RL). Pre-training grounds base LLMs in personal knowledge using corpora such as literature \cite{schwitzgebel2023creating} and personal descriptions \cite{salemi2024lamp}. Fine-tuning adapts models to target behaviors via instruction-style data \cite{shao2023character,ge2024scaling,sun2025identity}. RL refines behavior in interactive environments by optimizing reward signals that reflect response quality and persona consistency \cite{shea2023building,jang2023personalized,park2024diminished}.


For all individual-level simulation methods, they remain grounded in ``behaviorism", with limited exploration of deeper psychological motivations and underlying causes \cite{li2025simulating}. Besides, these methods struggle to ensure consistency and stability over long periods of simulation, often diverging from actual human decision-making behavior.

\vspace{-0.5em}
\subsubsection{Collective-level Simulation}
\vspace{-0.2em}

Mirroring the interconnected nature of human society, collective-level simulation targets both inter-agent interactions and their dynamics within a pre-defined context. We group these methods into two categories by environment complexity and agent diversity.


\textbf{Structured Scenario Interactions} place agents in controlled settings to study coordination and interaction protocols. \emph{Dialogue-centred} settings task agents with practising social interaction \cite{zhou2024sotopia,wang2024towards}, collaborative reasoning \cite{du2023improving}, or strategic communication in games \cite{xu2023exploring,wang2023avalon}. \emph{Goal-centred} settings assign agents specialised roles to complete domain tasks such as software engineering \cite{hong2023metagpt,qian2024chatdev}, clinical diagnosis \cite{li2024agent}, and judicial decision-making \cite{he2024agentscourt}.


\textbf{Emergent Societal Dynamics} focuses on more heterogeneous agent populations to reproduce non-linear interactions and system-level outcomes that are difficult to anticipate from individual behavior alone \cite{schelling1971dynamic}. 
For example, in general economics, simulations range from strategic-interaction settings (\eg, repeated games such as the Prisoner’s Dilemma \cite{xie2024can,mozikov2024good}) to macroeconomic simulations that model economy-wide dynamics and aggregate trends \cite{li2024econagent}.
In sociology and political science, simulations are used to approximate survey respondents and interviewees for opinion polling, election forecasting, and public-administration stress tests \cite{argyle2023out,chaudhary2024large,zhang2024electionsim}, and to probe theory-driven questions in controlled social settings \cite{chuang2024wisdom}. Online-platform simulations further model opinion dynamics on social media and related digital environments \cite{liu2024skepticism}.


\vspace{-1em}
\section{Fundamental Mismatches and Gaps}
\vspace{-0.5em}

\label{sec:limitations}

\subsection{Fundamental Mismatches}
\vspace{-0.3em}

We hereby use \emph{mismatch} to denote a systematic gap between \textbf{(1)} what LLM-integrated agent pipelines are typically optimized and validated to achieve, and \textbf{(2)} what social simulation ultimately demands to support credible population-level conclusions. In much current practice, the implicit objective is \emph{behavioral plausibility}: agents should produce coherent, role-consistent, and contextually appropriate text, often judged via transcripts, short-horizon consistency checks, or human preference judgments \cite{park2023generative,gao2024large,piao2025agentsociety}. By contrast, simulation-as-science requires \emph{mechanistic and counterfactual reliability}: outcomes should be robust to irrelevant prompt details, sensitive to the right causal interventions, and interpretable as consequences of well-specified assumptions rather than artifacts of the prompting/implementation stack.

This distinction matters because social simulations are frequently used not only to ``tell a compelling story'' but to support comparative claims (\eg, \emph{intervention A} versus \emph{intervention B}), to probe failure modes, or to explore policies that cannot be tested in the real world. If the validation target is local plausibility while the intended inference is causal or counterfactual, then seemingly minor design choices (\eg, persona wording, memory truncation, interaction protocol, or tool availability) can quietly dominate results \cite{park2023generative,ju2024sense,yang2024oasis}. In these cases, the simulation may be \emph{internally consistent} yet \emph{externally misleading}, producing conclusions that are brittle, irreproducible, or overconfident.

Framing issues as mismatches is thus a methodological move: it makes explicit where the community's proxy objectives (``does the agent look human?'') diverge from the epistemic goal (``can we trust the implied mechanism and the resulting comparative conclusions?''). The rest of this section uses this lens to pinpoint recurring mismatches that arise even when agents appear plausible, thereby motivating the evaluation and modeling gaps discussed next.

\textbf{Mismatch 1.} \emph{Role-playing plausibility does not imply faithful human simulation}.

A central promise of LLM-based social simulation is that, once an agent can ``sound like'' a target role (\eg, a voter, a policymaker, a patient), it can serve as a proxy for how a real human in that role would think and act. This is a mismatch because role-playing methods primarily optimize for \emph{linguistic and narrative coherence}, \eg, staying in-character, producing sensible justifications, and avoiding obvious contradictions, whereas faithful human simulation requires \emph{behavioral validity under constraints}: decisions should reflect stable preferences, bounded cognition, incentives, social learning, and norm compliance across time and across counterfactual interventions \cite{anthis2025position}. In other words, an agent can produce a convincing transcript while still instantiating the wrong decision process.


This mismatch has two consequences. First, it creates an illusion of validity: ``human-like'' transcripts can lead researchers to over-trust the simulated mechanism and treat subjective plausibility as evidence. Second, it induces fragility and hidden confounding: because behavior is prompt- and interface-mediated, small changes in persona wording, memory format, tools, or interaction protocol can shift conclusions later interpreted as substantive policy or social-mechanism effects \cite{park2023generative,ju2024sense,yang2024oasis}. For comparative claims (\eg, messaging strategies, institutional rules), this yields spurious treatment effects, unstable recommendations, and misleading ``explanations'' of emergent outcomes. \emph{Neither} of the dominant individual-level approaches in Section \ref{sec:current} fully resolves this mismatch.

Specifically, prompt-based methods generally improve \emph{surface role adherence} by conditioning the model on demographic descriptors, backstories, or exemplars. However, they typically treat the persona as a static textual constraint and rely on the base model to fill in missing structure. We argue that this makes them vulnerable to \textbf{(1)} \emph{underspecification} (\ie, many different latent decision rules can yield equally plausible text), \textbf{(2)} \emph{stereotype completion} (\ie, agents default to socially salient but inaccurate heuristics), and \textbf{(3)} \emph{prompt sensitivity} (\eg, behavior shifts with phrasing, ordering). Accordingly, prompt-based agents may be consistent in the short term while drifting in long-horizon interaction or responding unrealistically to incentives, uncertainty, and social pressure. Training-based methods (\eg, fine-tuning) can make agents more consistent and can better imitate a target style or population. Yet, the typical training signals still reward \emph{predictive or preference-aligned outputs}, not faithful decision-making under explicit institutional and informational constraints. Fine-tuning on narrative data may improve how an agent \emph{describes} motivations without ensuring that it \emph{acts} according to them when incentives change. Moreover, training-based agents inherit dataset biases and may generalize poorly under distribution shift \cite{sharma2024towards}, which is precisely the regime of interest in counterfactual social simulation. In short, stronger imitation does not automatically produce the causal fidelity required for trustworthy simulation.

\textbf{Mismatch 2.} \emph{Social simulation cannot be simply reduced to agent–agent interaction}.

The second mismatch arises from an interaction-centric view of ``society,'' where researchers build a network of agents, let them exchange messages, and then interpret the resulting conversation dynamics as a proxy for real-world collective behavior. This is a mismatch because, for humans, many consequential behaviors are mediated less by what others \emph{say} and more by what the world \emph{makes possible} and \emph{makes salient}: the opportunities, constraints, frictions, and incentives created by institutions and platforms. Accordingly, many social outcomes are not generated by dialogue alone; they are produced by \emph{agent--environment co-dynamics} \cite{schelling1971dynamic,bruch2015agent,koster2022spurious}. Here, ``environment'' includes not only physical constraints, but also institutional rules, market and platform mechanisms, information exposure and ranking systems, resource constraints, enforcement, and feedback loops that shape what agents can observe, what actions are available, and which actions are rewarded or punished.

Reducing social simulation to agent--agent interaction implicitly assumes that the environment is either neutral (a passive stage) or can be approximated by a simple message-passing protocol. In real settings, however, the environment is often the dominant causal channel. For example, the same population can exhibit qualitatively different collective outcomes under different moderation policies, recommendation algorithms, voting rules, pricing mechanisms, or enforcement regimes \cite{yang2024oasis}, even if individuals' ``intrinsic'' preferences remain unchanged. When such mechanisms are missing or overly simplified, simulations risk attributing emergent effects to agents' ``social behavior'' while the true driver in the world is the mediation layer.

The consequences are particularly severe when simulations are used to evaluate interventions. If an intervention in the real world operates by changing exposure, incentives, or institutional constraints (\eg, throttling misinformation, changing audit probability, altering default options, adjusting payoffs, modifying group formation), then an interaction-only simulation may be unable to represent the intervention at all, or will represent it as an ad hoc prompt instruction. This can yield false negatives (\ie, missing real effects), false positives (\ie, hallucinated effects), and incorrect mechanism stories that do not transfer outside the simulator. More broadly, it encourages a misleading decomposition of social phenomena into ``micro dialogue quality'' and ``macro emergence,'' obscuring the fact that macro patterns are often artifacts of the environment model.


\vspace{-1em}
\subsection{Current Gaps}
\vspace{-0.5em}

The two mismatches above point to paradigmatic flaws: conflating role-playing with human fidelity, and conflating social dynamics with inter-agent messaging while under-modeling the mediating role of the environment. Beyond these conceptual issues, the current literature also exhibits technical and workflow-level \emph{gaps}. We use \emph{gap} to denote a concrete yet underdeveloped or missing component in the end-to-end simulation pipeline that is necessary to \textbf{(1)} make the simulator's underlying assumptions explicit and empirically testable, and \textbf{(2)} evaluate whether an observed collective outcome remains robust beyond a particular choice of prompt or implementation detail. Unlike \emph{mismatches}, which concern discrepancies between what researchers claim and what their simulator can epistemically support, \emph{gaps} concern parts of the pipeline that remain inadequately operationalized.


\vspace{-0.6em}
\subsubsection{Evaluation of Simulation Results}
\vspace{-0.4em}

A central barrier to treating LLM-based agent societies as credible social simulators is the current lack of rigorous and comprehensive evaluation protocols. Much of the literature relies on demonstration-style validation: a small number of curated transcripts or trajectories are presented as evidence of ``realism'', with limited quantification of run-to-run variance and limited comparison against explicit baselines. This issue is particularly acute at the collective level, where outcomes are high-dimensional and real-world ``ground-truth'' trajectories are rarely available.

Across both the individual and collective levels, a small number of relatively more objective evaluation methods have begun to emerge; nevertheless, evaluation remains fragmented and far from standardized. At the individual level, existing protocols are relatively more structured and can be broadly grouped into \emph{static} and \emph{interactive} regimes \cite{mou2024individual}. Specifically, \textbf{static subjective evaluation} uses human or LLM judges to rate persona fidelity, coherence, and psychological plausibility, often via interview-style prompts \cite{gao2023livechat,wang2024rolellm,wang2024incharacter}; \textbf{Static objective evaluation} translates persona fidelity into constrained tasks with measurable targets, such as psychometric tests or multiple-choice questionnaires \cite{gao2023livechat,li2024steerability,ahn2023mpchat}. A complementary objective line adopts reference-based text similarity metrics in character-aligned dialogue or personalization benchmarks, reporting BLEU/ROUGE-style scores (or closely related overlap measures) to quantify alignment with target utterances and styles \cite{liu2022improving,chen2023large,gao2023livechat}; Finally, \textbf{interactive evaluation} embeds the agent in a multi-turn environment and measures performance via task/game outcomes under explicit rules \cite{shao2023character,chawla2023selfish,light2023avalonbench}.

Despite these efforts, most existing evaluations remain fundamentally output-centric and behaviorist. They validate agents through observable text, and treat fluency, stylistic consistency, and short-horizon coherence as proxies for human realism. Interview-based ratings, for instance, can be dominated by narrative plausibility rather than by whether the agent’s underlying preferences and constraints are faithfully instantiated \cite{piao2025agentsociety}. Reference-based text metrics primarily reward surface-level lexical alignment, and thus provide only a coarse proxy for open-ended social behavior and decision validity under incentives or information constraints \cite{nainia2025beyond}. Consequently, an agent may score well while still exhibiting unstable beliefs, prompt-induced shifts in preferences, or implausible adaptation under counterfactual interventions. This misalignment between what is measured and what is required for credible simulation constitutes a major technical evaluation gap.

At the collective level, evaluation is even less standardized and is typically framed in two complementary ways. In sandbox settings without reliable real-world traces, authors often ask whether the simulated society ``looks human-like,'' relying on Turing-test-style judgments of interaction quality or the plausibility of emergent narratives, as in early sandbox society simulations \cite{park2023generative}. In policy- or platform-facing settings where empirical traces exist, evaluation is more often cast as \emph{alignment}: does the simulator reproduce observed aggregate patterns, such as diffusion scale, opinion distributions, and temporal trends \cite{yang2024oasis,huang2026dualmind}. 

These evaluation practices also introduce several failure modes that directly threaten the epistemic value of social simulation. Firstly, they can enable selective reporting and confirmation bias: when only a few trajectories are highlighted, it becomes easy to overlook instability across random seeds, prompt variants, or sampling settings. Secondly, they can make method comparison ill-posed, because improvements in reported metrics may reflect better linguistic polish rather than improved behavioral validity. Thirdly, when collective simulations are used for decision support (\eg, public opinion diffusion or platform governance), weak validation can translate into brittle or biased recommendations, especially under distribution shift or information asymmetry. Overall, the current evaluation gap amplifies the risk that agent-based simulators are interpreted as causal or policy-relevant models, even when they have only been assessed as plausible generators of text.

\subsubsection{Interaction Dynamics and State Update}

Even assuming high-fidelity individual agents, a social simulator is well-defined only once it specifies an \emph{interaction kernel}: who can affect whom, through what channels, and how these interactions update persistent states (\eg, beliefs, relationships, resources, exposure). In current AI agent-based simulations, these mechanisms are often left implicit because the simulator interfaces are text-based: even in rich sandbox environments, world state and institutional rules are commonly represented via text, with prompt engineering acting as the de facto ``dynamics layer'' \cite{gao2024large}.


Existing work consequently implements inter-agent interaction in two broad styles. \emph{Dialogue-first} designs place agents in a shared conversational loop with unstructured natural language as the interaction medium, typical of sandbox ``society'' settings such as Generative Agents \cite{park2023generative}. \emph{Rule-mediated} designs couple agents through an explicit rule layer: agents communicate via structured formats (\eg, code/JSON) or act in environments with discrete action spaces and state-transition rules, so that outputs map to well-defined state changes \cite{mou2024individual}. Even within text-centric systems, interaction can be \emph{direct} (\ie, text exchange) or \emph{indirect} (\eg, coupling through predefined system rules), implying distinct causal pathways for how ``interaction'' propagates \cite{gao2024large}.


The dialogue-first approach is attractive because it maximizes expressiveness and reduces up-front modeling: in principle, any social act can be expressed as language. However, this flexibility comes with underspecified semantics. Unstructured language is inherently ambiguous and redundant \cite{mou2024individual}, so utterances may mix narration, intention, and action, and the same ``interaction'' can take many surface forms. This blurs the boundary between \emph{communication} and \emph{state transition}, making macro trajectories sensitive to prompt phrasing, verbosity, and turn-taking order rather than stable behavioral mechanisms. Moreover, when interaction effects must flow primarily through a shared conversational channel, the simulator implicitly assumes synchronous and highly observable exchanges, misaligned with real social systems involving parallel, partially observed interactions across time scales.


Rule-mediated designs attempt to restore well-posed interaction by explicitly separating agent decisions from environment updates. In practice, this includes ``hard'' rules such as diffusion/contagion processes: for example, S3 simulates individuals embedded in a social network to study information diffusion and the dynamics of emotions and attitudes \cite{gao2023s3}, and epidemic-oriented simulations test whether LLM agents can reproduce complex phenomena such as multi-peak infection patterns \cite{williams2023epidemic}. Platform simulators such as OASIS further operationalize interaction through an environment server and recommender-mediated exposure, where a time engine activates agents and an action module updates environment state (\eg, new posts and evolving relations) \cite{yang2024oasis}. While this rule layer improves clarity and reproducibility, it also reintroduces a classical ABM tension: rules and models cannot cover all dimensions of real-world heterogeneity, and using rules to drive behavior may only capture limited aspects of diverse preferences and decision processes \cite{gao2024large}. When the rule layer dominates, LLM agents risk becoming narrative wrappers around a largely pre-specified dynamics, weakening the simulator’s ability to generate new causal insight.

Crucially, even after choosing an interaction style, the simulator must still specify \emph{state update}: which interactions happen concurrently, how they are interleaved, and what each agent observes when deciding. Many implementations execute interactions as sequential loops (fixed order or batched), yet real societies exhibit asynchronous and heterogeneous activity: agents can receive multiple signals between decisions, interact through different channels, and update different state variables on different time scales. AgentSociety explicitly notes that common multi-agent frameworks rely on predefined message-passing procedures to determine execution sequences, whereas real-world decisions are not strictly dependent on other agents’ immediate inputs, making asynchronous societal dynamics hard to simulate \cite{piao2025agentsociety}. In parallel, OASIS emphasizes temporal activation (via a time engine) and real-time environment updates after actions \cite{yang2024oasis}. Together, these examples suggest that scheduling and synchronization are not mere engineering choices but part of the model: changing them can alter exposure, feedback loops, and emergent outcomes.

\vspace{-0.5em}
\subsubsection{Initialization and Information Priors}

A third (and often underestimated) barrier to credible LLM-based social simulation is the under-specification of \emph{initial conditions} and \emph{information priors}. Even with highly role-faithful agents and a well-defined state-update rule, a simulator must still decide \emph{who exists}, \emph{who is connected to whom}, and \emph{who knows what at $t=0$}. In policy- and event-facing settings, these choices are not merely structural: they determine how a new shock (\eg, a regulation, crisis, or public announcement) is instantiated as an initial perturbation of the system, and therefore whether ``short-run'' effects are even well-defined or interpretable. In much of current practice, however, initialization is treated primarily as an engineering convenience, \eg, sampling $N$ personas, wiring an interaction graph, or embedding a scenario description directly into agent prompts \cite{yang2024oasis}. Such procedures implicitly fix strong assumptions about the initial distribution of awareness, exposure, and belief, effectively hard-coding the immediate impact of the shock rather than modeling its emergence. As a result, these unexamined information priors can dominate downstream dynamics, making simulation outcomes difficult to interpret and epistemically unreliable in practice.

\noindent \textbf{Role and Population Assignment Involves More than Matching Population-level Statistics.} A common starting point is to assign each agent a role or persona by matching coarse population statistics (\eg, age, gender, occupation, or ideology) and then sampling individuals accordingly. This approach is attractive because it requires only aggregate distributions and scales easily; it also aligns with the ``silicon samples'' framing, in which LLMs are used to elicit population-level responses from demographically conditioned prompts \cite{argyle2023out}. Recent large-scale simulations adopt similar strategies by constructing personas from survey instruments and census-like sources \cite{tornberg2023simulating,yang2024oasis,zhang2024electionsim}. However, matching population statistics is rarely sufficient for credible social simulation, because social behavior depends on structured heterogeneity, \eg, correlations among attributes, institutional positions, and latent variables that are not captured by a short persona description. Naively sampling attributes independently can break the underlying joint distribution, produce implausible combinations, and distort which agents become influential or representative in downstream dynamics. Some systems explicitly acknowledge this limitation by employing techniques such as iterative proportional fitting to approximate joint demographic distributions from multiple marginal constraints \cite{zhang2024electionsim}. Yet even when demographic joints are matched, the mapping from demographics to \emph{beliefs, incentives, and constraints} remains fundamentally underdetermined. Consequently, two simulation initializations may share identical aggregate statistics while encoding very different mechanisms of persuasion, norm compliance, or strategic adaptation, precisely the mechanisms that policy-oriented social simulations seek to interrogate.

\noindent \textbf{Scale Selection is an Epistemic Modeling Choice, Not Merely a Computational Constraint.} A second design question concerns how many agents to instantiate. The principled choice, simulating the full target population, is typically infeasible for LLM-driven agents due to cost and latency. Most existing social simulations therefore operate at substantially reduced scale: Tornberg \etal use 500 agents to study algorithmic news-feed interventions \cite{tornberg2023simulating}; HiSim samples 1,000 participating users per movement dataset and models 300 as ``core'' LLM-driven users \cite{mou2024unveiling}; and ElectionSim processes over one million distinct social media users to construct a million-level voter pool \cite{zhang2024electionsim}. Large by LLM-agent standards, these scales nevertheless remain orders of magnitude smaller than the populations implicated in national elections, platform governance, or public health policy. Down-sampling ``by proportion'' is not innocuous: many collective phenomena exhibit strong non-linear dependencies on population size and interaction topology. Rare subgroups can trigger cascades; heavy-tailed influence structures can cause outcomes to hinge on a few high-degree nodes; and finite-size effects can inflate run-to-run variance, producing apparent ``treatment effects'' that vanish at larger scales. HiSim's hybrid design, combining core LLM-driven agents with a larger deductive pool, is an explicit attempt to navigate this precision-scale trade-off \cite{mou2024unveiling}. Yet the field lacks principled guidance on when reduced-scale simulations preserve the relevant invariants of the underlying social system, and when they instead introduce qualitative artifacts that compromise inference in practice.

\noindent \textbf{Conflating Information Availability with Information Awareness.} 
Beyond \emph{who} is simulated, any agent-based simulation must also specify \emph{what each agent initially knows and believes}, a design choice that is particularly consequential in policy-oriented applications, where conclusions often hinge on how information, regulations, or announcements are perceived and acted upon. In contemporary LLM-agent pipelines, this epistemic initialization is most often implemented through prompt-level priors, including a persona description, a brief memory summary, and an explicit textual specification of the scenario and intervention. In HiSim, for instance, real-world trigger events are introduced as natural-language ``background information'' provided to core users \cite{mou2024unveiling}. Similarly, in fake-news simulations, new information is typically operationalized as discrete text content that is broadcast by designated modules or ``infected'' individuals \cite{liu2024skepticism,huang2026dualmind}. While such mechanisms are understandable, as a simulator must introduce exogenous shocks in some form, they implicitly collapse a critical distinction between \emph{the existence of information} and \emph{agents' awareness of that information}.


\vspace{-0.8em}
\section{Actions for Future Social Simulation}
\vspace{-0.5em}


Many of the issues discussed above arise from the lack of a comprehensive and standardized formulation of AI agent-based social simulation. To address this gap, we propose a unified formulation that makes the structural requirements for credible simulation explicit, and use it to derive concrete actions for improving the design, evaluation, and interpretability of agent-based social simulations.

\begin{definition}[Environment-involved AI Agent-based Social Simulator]
\label{def:simulator}
For a fixed number of agents $N$ and simulation horizon $T$, the simulator is defined as a tuple 
\begin{equation}
\texttt{Sim} \triangleq \Bigl(\mathcal{E}, \mathcal{G}, \mathcal{C}, \{\mathcal{M}_i, \mathcal{O}_i, \mathcal{A}_i, \mathcal{P}_i, \mathcal{U}_i, \mathcal{R}_i\}_{i=1}^N, D_0, \texttt{Sch}, \texttt{Vis}, \texttt{Tr} \Bigr),
\end{equation}
where
$\mathcal{E}$ is the \textbf{Environment} state space, with $e_t \in \mathcal{E}$ denoting the instantiated configuration of the social context at time $t$, including policy parameters, institutional settings, resource levels, and enforcement intensities; $\mathcal{G}$ is the \textbf{Graph} (network) state space, with $g_t \in \mathcal{G}$; $\mathcal{C}$ is the \textbf{Context} space, with $c_t \in \mathcal{C}$ showing exogenous conditions (\eg, policies, events, or scenario settings); For each agent $i$, $\mathcal{M}_i$ is its \textbf{Mental} state space with $m_{t}^i \in \mathcal{M}_i$ indicating information like beliefs, preferences, memory, $\mathcal{O}_i$ is its \textbf{Observation} space with $o_{t}^i \in \mathcal{O}_i$, $\mathcal{A}_i$ is its \textbf{Action} space with $a_{t}^i \in \mathcal{A}_i$. $P_i$ is the (LLM-based) \textbf{Policy}, $s.t.$ $a_t^i \sim P_i(\cdot|o_t^i, m_t^i; \ell_i)$, where $\ell_i$ is the LLM configuration (\eg, checkpoint, prompting template, tool access). $U_i$ is the \textbf{Update} function for the mental state \ie, $m_{t+1}^i = U_i(m_{t}^i, o_{t}^i, a_{t}^i, v_t, e_t, g_t, c_t)$. Each $\mathcal{R}_i$ is a \textbf{Reward} function, formalizing how institutional incentives, costs, and enforcement mechanisms translate agent actions into social consequences, thereby shaping the agent–environment co-dynamics; $D_0$ is the \textbf{Distribution} over initial conditions: $(e_0, g_0, m_0^{1:N}) \sim D_0(\cdot|c_0)$. It must encode population composition, attribute correlations, initial network structure, and epistemic priors; $\texttt{Sch}$ is the \textbf{Scheduler} returning the active index set: $I_t = \texttt{Sch}(e_t, g_t, c_t) \subseteq \{1, \cdots, N\}$, modeling choice that formalizes when and which agents are allowed to act; $\texttt{Vis}$ is the \textbf{Visibility} mechanism, $s.t.$ $(o_t^{1:N}, v_t) \sim \texttt{Vis}(\cdot|e_t, g_t, c_t)$, where $v_t$ is a visibility object (\eg, exposure graph) that determines who sees what and thus represents information asymmetries; $\texttt{Tr}$ is the \textbf{Transition} for environment and network dynamics: $(e_{t+1}, g_{t+1}) \sim \texttt{Tr}(\cdot
|e_t,g_t,c_t, a_t^{1:N})$. Specifically, given $c_{0:T}$ and an initial draw from $D_0$, the simulator proceeds for $t=1, \cdots, T$ as follows: \textbf{(1)} sample $(o_t^{1:N}, v_t)$ via $\texttt{Vis}$; \textbf{(2)} sample active set $I_t$ via $\texttt{Sch}$; \textbf{(3)} for each $i \in I_t$, sample $a_t^i$ via $P_i$ (and optionally set $a_t^i$ to a no-op for $i \notin I_t$); \textbf{(4)} update $\{m_{t+1}^i\}$ via $\{U_i\}$; \textbf{(5)} transition $(e_{t+1, g_{t+1}})$ via $\texttt{Tr}$. Specific instantiations are application-dependent and beyond our scope.
\end{definition}

\vspace{-0.5em}

This formulation potentially addresses the mismatches and gaps in Section \ref{sec:limitations} by making previously implicit assumptions explicit and controllable. By separating agent-level generation from environment-mediated mechanisms, the simulator distinguishes how information is exposed, when agents may act, and how actions produce institutional consequences. Specifically, information asymmetries are represented through the visibility mechanism rather than by assuming globally shared context; interaction timing and ordering are governed by the scheduler, preventing turn-taking artifacts from being conflated with social dynamics; and incentives and enforcement are encoded via consequence functions, allowing policy interventions to be modeled as controlled changes to environment state and mechanisms rather than prompt-only modifications. While this formulation \emph{does not guarantee valid social simulation}, it provides a \emph{structured interface} for evaluating robustness, sensitivity, and counterfactual reliability.


Based on the aforementioned formulation and understandings, we advocate the following actions:


\noindent \textbf{Action 1.} \emph{Treat the environment as a first-class, auditable object}. A credible agent-based social simulator should make the environment explicit, rather than embedding it in natural-language prompts or ad hoc implementation details. Concretely, the environment state $e_t$ and its governing mechanisms, including $\texttt{Vis}$ (exposure and information routing), $\texttt{Sch}$ (action scheduling), $\texttt{Tr}$ (environment dynamics), and the incentive mapping $R$, should be specified as versioned, inspectable artifacts with logged traces, enabling external auditors to reconstruct the constraints, incentives, and information frictions at each step. This explicitness directly reduces hidden confounding: macro-level outcomes trace to declared mechanism choices rather than prompt phrasing or undocumented simulator defaults. When manual specification of $e_t$ and $\texttt{Tr}$ is infeasible, one can learn an explicit dynamics model from empirical trajectories (\eg, a ``world model'' \cite{zhou2025social}), provided its intermediate states and uncertainty diagnostics remain exposed, keeping the environment auditable rather than subsumed into the LLM.

\noindent \textbf{Action 2.} \emph{Move evaluation beyond plausibility to mechanistic and counterfactual reliability}. Current practice often evaluates social simulations by transcript plausibility or short-horizon coherence, yet such criteria are misaligned with the epistemic goal of drawing mechanism- or policy-relevant conclusions. Evaluation should instead directly probe mechanistic and counterfactual reliability: \textbf{(1)} whether agent decisions respond appropriately to the constraints and incentives encoded in $e_t$ and $R$ (behavior under constraints); \textbf{(2)} whether targeted edits to $\mathcal{C}$, $\texttt{Vis}$, $\texttt{Sch}$, or $\texttt{Tr}$ induce stable and directionally consistent changes in macro-level outcomes (mechanism sensitivity); and \textbf{(3)} whether comparative conclusions persist across random seeds, prompt paraphrases, and other nominally irrelevant implementation choices (counterfactual stability). These criteria can be operationalized through mechanism ablations, negative controls, and counterfactual sweeps that vary one component at a time, with results reported as distributions (\eg, uncertainty intervals and variance decomposition) rather than selected trajectories. This reframes evaluation from ``does it look human-like?'' to ``can the implied mechanism and comparative claims be trusted?''.

\noindent \textbf{Action 3.} \emph{Interpret simulation outcomes with epistemic caution and explicit uncertainty}. Finally, and most importantly, simulation outputs should be interpreted with epistemic caution: even under an explicit formulation, they do not constitute automatic evidence of social dynamics, but only conditional implications of the assumed mechanisms and implementation choices. We therefore recommend treating uncertainty as a first-class output and reporting sensitivity to $D_0$, $\texttt{Vis}$, $\texttt{Sch}$, $\texttt{Tr}$, and the LLM configuration $\ell_i$. Without such uncertainty reporting, mechanistic or policy-facing conclusions risk being non-transferable and overconfident.





\vspace{-1em}
\section{Alternative Views}
\vspace{-0.8em}

A first credible counter-view, which we term the \emph{individual-fidelity-sufficient thesis}, holds that the main bottleneck in AI agent-based social simulation is the fidelity of individual agents, and that sufficiently human-aligned agents would naturally yield realistic collective behaviour, rendering explicit environment modeling secondary. This view is articulated by \cite{anthis2025position} and reinforced by demonstrations of LLMs reproducing survey responses and electoral preferences when conditioned on demographic personas \cite{argyle2023out, park2023generative, zhang2024electionsim}; a parallel constructive line proposes deepening the cognitive substrate of agents \cite{li2025simulating}. We respect this view but disagree with its implication that environment modeling is secondary. Methods that refine an agent's internal generation or decision heuristics do not specify how information is produced or disclosed, nor how incentives and constraints are implemented and enforced. Consequently, even highly role-consistent agents placed in an unrealistic or underspecified environment can generate collective dynamics that are internally coherent yet externally implausible \cite{ju2024sense, yang2024oasis}. Treating the environment as an afterthought is therefore an epistemic error: without explicitly modeling how agents are coupled to their world, making agents more human-like does not by itself make the simulation more faithful.


A second concern, the \emph{rigidity worry}, holds that explicit environment specification may be infeasible or counterproductive: formalising environment dynamics or institutional rules is sometimes seen as reintroducing the rigid assumptions of classical rule-based models \cite{wolfram1984cellular, gatti2011macroeconomics, lengnick2013agent}, reducing LLM agents to narrative wrappers around fixed mechanisms. We acknowledge this tension. However, removing explicit structure does not eliminate assumptions; it relocates them into prompts, scheduling, and interfaces, where they become implicit, brittle, and hard to audit. Our position is not that environments must be fully specified, but that making key mechanisms explicit, even approximately, enables stress-testing, comparison, and diagnosis. Without such framework, it becomes difficult to attribute outcomes to underlying mechanisms, limiting the scientific value of agent-based social simulation.


\vspace{-1em}
\section{Conclusion}
\vspace{-0.8em}

Large language models have expanded agent-based social simulation, but expressiveness alone does not confer epistemic validity. We argued that AI agents alone are not (yet) sufficient for social simulation: the central challenge is not solely how human-like agents appear, but whether the assumptions governing interaction, information exposure, and institutional coupling are made explicit, testable, and auditable. Without such structure, simulations risk becoming persuasive narratives rather than scientific instruments. Our position is forward-looking rather than dismissive: we argue not against AI agents, but for a shift in how social simulations are framed, constructed, and evaluated. We call on the community to treat aggregate human-correlation as insufficient evidence, and mechanism-level audits as the default. Treating environments as first-class objects, separating proxy objectives from epistemic goals, and foregrounding uncertainty are necessary to move from demonstration to explanation, and from plausibility to inference. Only by embracing these principles can agent-based social simulation mature into a reliable tool for scientific understanding and policy reasoning, rather than a source of overconfident yet fragile conclusions.

\bibliographystyle{plain}
\bibliography{references}

@article{zhang2026agents,
  title={Agents in the wild: Safety, society, and the illusion of sociality on moltbook},
  author={Zhang, Yunbei and Mei, Kai and Liu, Ming and Wang, Janet and Metaxas, Dimitris N and Wang, Xiao and Hamm, Jihun and Ge, Yingqiang},
  journal={arXiv preprint arXiv:2602.13284},
  year={2026}
}

@book{gilbert2005simulation,
  title={Simulation for the social scientist},
  author={Gilbert, Nigel and Troitzsch, Klaus},
  year={2005},
  publisher={McGraw-Hill Education (UK)}
}

@book{devaus2013surveys,
  author    = {{de Vaus}, David},
  title     = {Surveys in Social Research},
  edition   = {6},
  year      = {2013},
  publisher = {Routledge},
  doi       = {10.4324/9780203519196},
  url       = {https://doi.org/10.4324/9780203519196}
}

@article{falk2009lab,
  title={Lab experiments are a major source of knowledge in the social sciences},
  author={Falk, Armin and Heckman, James J},
  journal={Science},
  volume={326},
  number={5952},
  pages={535--538},
  year={2009},
  publisher={American Association for the Advancement of Science}
}

@article{baldassarri2017field,
  title={Field experiments across the social sciences},
  author={Baldassarri, Delia and Abascal, Maria},
  journal={Annual review of sociology},
  volume={43},
  number={1},
  pages={41--73},
  year={2017},
  publisher={Annual Reviews}
}

@article{bruch2015agent,
  title={Agent-based models in empirical social research},
  author={Bruch, Elizabeth and Atwell, Jon},
  journal={Sociological methods \& research},
  volume={44},
  number={2},
  pages={186--221},
  year={2015},
  publisher={SAGE Publications Sage CA: Los Angeles, CA}
}

@article{faucher2022agent,
  title={Agent-based modelling of reactive vaccination of workplaces and schools against COVID-19},
  author={Faucher, Benjamin and Assab, Rania and Roux, Jonathan and Levy-Bruhl, Daniel and Tran Kiem, C{\'e}cile and Cauchemez, Simon and Zanetti, Laura and Colizza, Vittoria and Bo{\"e}lle, Pierre-Yves and Poletto, Chiara},
  journal={Nature communications},
  volume={13},
  number={1},
  pages={1414},
  year={2022},
  publisher={Nature Publishing Group UK London}
}

@article{li2020opinion,
  title={Opinion dynamics model based on the cognitive dissonance: An agent-based simulation},
  author={Li, Ke and Liang, Haiming and Kou, Gang and Dong, Yucheng},
  journal={Information Fusion},
  volume={56},
  pages={1--14},
  year={2020},
  publisher={Elsevier}
}

@article{chen2012agent,
  title={Agent-based modeling of the effects of social norms on enrollment in payments for ecosystem services},
  author={Chen, Xiaodong and Lupi, Frank and An, Li and Sheely, Ryan and Vi{\~n}a, Andr{\'e}s and Liu, Jianguo},
  journal={Ecological modelling},
  volume={229},
  pages={16--24},
  year={2012},
  publisher={Elsevier}
}

@article{granha2022opinion,
  title={Opinion dynamics in financial markets via random networks},
  author={Granha, Mateus FB and Vilela, Andr{\'e} LM and Wang, Chao and Nelson, Kenric P and Stanley, H Eugene},
  journal={Proceedings of the National Academy of Sciences},
  volume={119},
  number={49},
  pages={e2201573119},
  year={2022},
  publisher={National Academy of Sciences}
}

@article{wolfram1984cellular,
  title={Cellular automata as models of complexity},
  author={Wolfram, Stephen},
  journal={Nature},
  volume={311},
  number={5985},
  pages={419--424},
  year={1984},
  publisher={Nature Publishing Group UK London}
}

@article{lengnick2013agent,
  title={Agent-based macroeconomics: A baseline model},
  author={Lengnick, Matthias},
  journal={Journal of Economic Behavior \& Organization},
  volume={86},
  pages={102--120},
  year={2013},
  publisher={Elsevier}
}

@book{gatti2011macroeconomics,
  title={Macroeconomics from the Bottom-up},
  author={Gatti, Domenico Delli and Desiderio, Saul and Gaffeo, Edoardo and Cirillo, Pasquale and Gallegati, Mauro},
  volume={1},
  year={2011},
  publisher={Springer Science \& Business Media}
}

@article{vinkovic2006physical,
  title={A physical analogue of the Schelling model},
  author={Vinkovi{\'c}, Dejan and Kirman, Alan},
  journal={Proceedings of the National Academy of Sciences},
  volume={103},
  number={51},
  pages={19261--19265},
  year={2006},
  publisher={National Academy of Sciences}
}

@article{helbing1995social,
  title={Social force model for pedestrian dynamics},
  author = {Helbing, Dirk and Moln{\'a}r, P{\'e}ter},
  journal={Physical review E},
  volume={51},
  number={5},
  pages={4282},
  year={1995},
  publisher={APS}
}

@article{gao2024large,
  title={Large language models empowered agent-based modeling and simulation: A survey and perspectives},
  author={Gao, Chen and Lan, Xiaochong and Li, Nian and Yuan, Yuan and Ding, Jingtao and Zhou, Zhilun and Xu, Fengli and Li, Yong},
  journal={Humanities and Social Sciences Communications},
  volume={11},
  number={1},
  pages={1--24},
  year={2024},
  publisher={Palgrave}
}

@inproceedings{wang2024incharacter,
  title={Incharacter: Evaluating personality fidelity in role-playing agents through psychological interviews},
  author={Wang, Xintao and Xiao, Yunze and Huang, Jen-tse and Yuan, Siyu and Xu, Rui and Guo, Haoran and Tu, Quan and Fei, Yaying and Leng, Ziang and Wang, Wei and others},
  booktitle={ACL},
  pages={1840--1873},
  year={2024}
}

@article{mou2024individual,
  title={From individual to society: A survey on social simulation driven by large language model-based agents},
  author={Mou, Xinyi and Ding, Xuanwen and He, Qi and Wang, Liang and Liang, Jingcong and Zhang, Xinnong and Sun, Libo and Lin, Jiayu and Zhou, Jie and Huang, Xuanjing and others},
  journal={arXiv preprint arXiv:2412.03563},
  year={2024}
}

@article{piao2025agentsociety,
  title={Agentsociety: Large-scale simulation of llm-driven generative agents advances understanding of human behaviors and society},
  author={Piao, Jinghua and Yan, Yuwei and Zhang, Jun and Li, Nian and Yan, Junbo and Lan, Xiaochong and Lu, Zhihong and Zheng, Zhiheng and Wang, Jing Yi and Zhou, Di and others},
  journal={arXiv preprint arXiv:2502.08691},
  year={2025}
}

@article{yang2024oasis,
  title={Oasis: Open agent social interaction simulations with one million agents},
  author={Yang, Ziyi and Zhang, Zaibin and Zheng, Zirui and Jiang, Yuxian and Gan, Ziyue and Wang, Zhiyu and Ling, Zijian and Chen, Jinsong and Ma, Martz and Dong, Bowen and others},
  journal={arXiv preprint arXiv:2411.11581},
  year={2024}
}

@inproceedings{liu2024skepticism,
  title={From Skepticism to Acceptance: Simulating the Attitude Dynamics Toward Fake News},
  author={Liu, Yuhan and Chen, Xiuying and Zhang, Xiaoqing and Gao, Xing and Zhang, Ji and Yan, Rui},
  booktitle={IJCAI},
  pages={7886--7894},
  year={2024}
}

@article{zhou2025social,
  title={Social world models},
  author={Zhou, Xuhui and Liu, Jiarui and Yerukola, Akhila and Kim, Hyunwoo and Sap, Maarten},
  journal={arXiv preprint arXiv:2509.00559},
  year={2025}
}

@inproceedings{mou2024unveiling,
  title={Unveiling the truth and facilitating change: Towards agent-based large-scale social movement simulation},
  author={Mou, Xinyi and Wei, Zhongyu and Huang, Xuan-Jing},
  booktitle={{ACL} (Findings)},
  pages={4789--4809},
  year={2024}
}

@article{argyle2023out,
  title={Out of one, many: Using language models to simulate human samples},
  author={Argyle, Lisa P and Busby, Ethan C and Fulda, Nancy and Gubler, Joshua R and Rytting, Christopher and Wingate, David},
  journal={Political Analysis},
  volume={31},
  number={3},
  pages={337--351},
  year={2023},
  publisher={Cambridge University Press}
}

@article{tornberg2023simulating,
  title={Simulating social media using large language models to evaluate alternative news feed algorithms},
  author={T{\"o}rnberg, Petter and Valeeva, Diliara and Uitermark, Justus and Bail, Christopher},
  journal={arXiv preprint arXiv:2310.05984},
  year={2023}
}

@inproceedings{huang2026dualmind,
  title={DualMind: Towards Understanding Cognitive-Affective Cascades in Public Opinion Dissemination via Multi-Agent Simulation},
  author={Huang, Enhao and Pan, Tongtong and Zhang, Shuhuai and Jin, Qishu and Zhen, Liheng and Hu, Kaichun and Li, Yiming and Qin, Zhan and Ren, Kui},
  booktitle={{WWW} (Demo Paper)},
  year={2026},
  publisher={ACM}
}

@inproceedings{nainia2025beyond,
  title={Beyond BLEU: Ethical Risks of Misleading Evaluation in Domain-Specific QA with LLMs},
  author={Nainia, Ayoub and Vignes-Lebbe, R{\'e}gine and Mousannif, Hajar and Zahir, Jihad},
  booktitle={Proceedings of the First Workshop on Comparative Performance Evaluation: From Rules to Language Models},
  pages={77--86},
  year={2025}
}

@article{achiam2023gpt,
  title={Gpt-4 technical report},
  author={Achiam, Josh and Adler, Steven and Agarwal, Sandhini and Ahmad, Lama and Akkaya, Ilge and Aleman, Florencia Leoni and Almeida, Diogo and Altenschmidt, Janko and Altman, Sam and Anadkat, Shyamal and others},
  journal={arXiv preprint arXiv:2303.08774},
  year={2023}
}

@article{touvron2023llama,
  title={Llama 2: Open foundation and fine-tuned chat models},
  author={Touvron, Hugo and Martin, Louis and Stone, Kevin and Albert, Peter and Almahairi, Amjad and Babaei, Yasmine and Bashlykov, Nikolay and Batra, Soumya and Bhargava, Prajjwal and Bhosale, Shruti and others},
  journal={arXiv preprint arXiv:2307.09288},
  year={2023}
}

@article{guo2025deepseek,
  title={DeepSeek-R1 incentivizes reasoning in LLMs through reinforcement learning},
  author={Guo, Daya and Yang, Dejian and Zhang, Haowei and Song, Junxiao and Wang, Peiyi and Zhu, Qihao and Xu, Runxin and Zhang, Ruoyu and Ma, Shirong and Bi, Xiao and others},
  journal={Nature},
  volume={645},
  number={8081},
  pages={633--638},
  year={2025},
  publisher={Nature Publishing Group UK London}
}

@inproceedings{park2023generative,
  title={Generative agents: Interactive simulacra of human behavior},
  author={Park, Joon Sung and O'Brien, Joseph and Cai, Carrie Jun and Morris, Meredith Ringel and Liang, Percy and Bernstein, Michael S},
  booktitle={UIST},
  pages={1--22},
  year={2023}
}

@article{shinn2023reflexion,
  title={Reflexion: Language agents with verbal reinforcement learning},
  author={Shinn, Noah and Cassano, Federico and Gopinath, Ashwin and Narasimhan, Karthik and Yao, Shunyu},
  journal={NeurIPS},
  pages={8634--8652},
  year={2023}
}

@article{wang2023voyager,
  title={Voyager: An open-ended embodied agent with large language models},
  author={Wang, Guanzhi and Xie, Yuqi and Jiang, Yunfan and Mandlekar, Ajay and Xiao, Chaowei and Zhu, Yuke and Fan, Linxi and Anandkumar, Anima},
  journal={arXiv preprint arXiv:2305.16291},
  year={2023}
}

@article{johnson2019hidden,
  title={Hidden resilience and adaptive dynamics of the global online hate ecology},
  author={Johnson, Nicola F and Leahy, Rhys and Restrepo, N Johnson and Vel{\'a}squez, Nicholas and Zheng, Minzhang and Manrique, Pedro and Devkota, Prajwal and Wuchty, Stefan},
  journal={Nature},
  volume={573},
  number={7773},
  pages={261--265},
  year={2019},
  publisher={Nature Publishing Group UK London}
}

@article{jiang2023evaluating,
  title={Evaluating and inducing personality in pre-trained language models},
  author={Jiang, Guangyuan and Xu, Manjie and Zhu, Song-Chun and Han, Wenjuan and Zhang, Chi and Zhu, Yixin},
  journal={NeurIPS},
  pages={10622--10643},
  year={2023}
}

@inproceedings{bhandari2025evaluating,
  title={Evaluating personality traits in large language models: Insights from psychological questionnaires},
  author={Bhandari, Pranav and Naseem, Usman and Datta, Amitava and Fay, Nicolas and Nasim, Mehwish},
  booktitle={{WWW} (Short Paper)},
  pages={868--872},
  year={2025}
}

@article{serapio2025psychometric,
  title     = {A psychometric framework for evaluating and shaping personality traits in large language models},
  author    = {Serapio-Garc{\'\i}a, Gregory and Safdari, Mustafa and Crepy, Cl{\'e}ment and Sun, Luning and Fitz, Stephen and Romero, Peter and Abdulhai, Marwa and Faust, Aleksandra and Matari{\'c}, Maja},
  journal   = {Nature Machine Intelligence},
  volume    = {7},
  number    = {12},
  pages     = {1954--1968},
  year      = {2025}
}

@article{shanahan2023role,
  title={Role play with large language models},
  author={Shanahan, Murray and McDonell, Kyle and Reynolds, Laria},
  journal={Nature},
  volume={623},
  number={7987},
  pages={493--498},
  year={2023},
  publisher={Nature Publishing Group UK London}
}

@article{zou2023representation,
  title={Representation engineering: A top-down approach to ai transparency},
  author={Zou, Andy and Phan, Long and Chen, Sarah and Campbell, James and Guo, Phillip and Ren, Richard and Pan, Alexander and Yin, Xuwang and Mazeika, Mantas and Dombrowski, Ann-Kathrin and others},
  journal={arXiv preprint arXiv:2310.01405},
  year={2023}
}

@inproceedings{sharma2024towards,
  title={Towards Understanding Sycophancy in Language Models},
  author={Sharma, Mrinank and Tong, Meg and Korbak, Tomasz and Duvenaud, David and Askell, Amanda and Bowman, Samuel R and DURMUS, Esin and Hatfield-Dodds, Zac and Johnston, Scott R and Kravec, Shauna M and others},
  booktitle={ICLR},
  year={2024}
}

@article{boiko2023autonomous,
  title={Autonomous chemical research with large language models},
  author={Boiko, Daniil A and MacKnight, Robert and Kline, Ben and Gomes, Gabe},
  journal={Nature},
  volume={624},
  number={7992},
  pages={570--578},
  year={2023},
  publisher={Nature Publishing Group UK London}
}

@article{romera2024mathematical,
  title={Mathematical discoveries from program search with large language models},
  author={Romera-Paredes, Bernardino and Barekatain, Mohammadamin and Novikov, Alexander and Balog, Matej and Kumar, M Pawan and Dupont, Emilien and Ruiz, Francisco JR and Ellenberg, Jordan S and Wang, Pengming and Fawzi, Omar and others},
  journal={Nature},
  volume={625},
  number={7995},
  pages={468--475},
  year={2024},
  publisher={Nature Publishing Group UK London}
}

@article{m2024augmenting,
  title={Augmenting large language models with chemistry tools},
  author={M. Bran, Andres and Cox, Sam and Schilter, Oliver and Baldassari, Carlo and White, Andrew D and Schwaller, Philippe},
  journal={Nature Machine Intelligence},
  volume={6},
  number={5},
  pages={525--535},
  year={2024},
  publisher={Nature Publishing Group UK London}
}

@inproceedings{hong2023metagpt,
  title={MetaGPT: Meta programming for a multi-agent collaborative framework},
  author={Hong, Sirui and Zhuge, Mingchen and Chen, Jonathan and Zheng, Xiawu and Cheng, Yuheng and Wang, Jinlin and Zhang, Ceyao and Wang, Zili and Yau, Steven Ka Shing and Lin, Zijuan and others},
  booktitle={ICLR},
  year={2023}
}

@article{yang2024swe,
  title={Swe-agent: Agent-computer interfaces enable automated software engineering},
  author={Yang, John and Jimenez, Carlos E and Wettig, Alexander and Lieret, Kilian and Yao, Shunyu and Narasimhan, Karthik and Press, Ofir},
  journal={NeurIPS},
  pages={50528--50652},
  year={2024}
}

@inproceedings{qian2024chatdev,
  title={Chatdev: Communicative agents for software development},
  author={Qian, Chen and Liu, Wei and Liu, Hongzhang and Chen, Nuo and Dang, Yufan and Li, Jiahao and Yang, Cheng and Chen, Weize and Su, Yusheng and Cong, Xin and others},
  booktitle={ACL},
  pages={15174--15186},
  year={2024}
}

@inproceedings{yao2023react,
  title={React: Synergizing reasoning and acting in language models},
  author={Yao, Shunyu and Zhao, Jeffrey and Yu, Dian and Du, Nan and Shafran, Izhak and Narasimhan, Karthik R and Cao, Yuan},
  booktitle={ICLR},
  year={2023}
}

@article{shen2023hugginggpt,
  title={Hugginggpt: Solving ai tasks with chatgpt and its friends in hugging face},
  author={Shen, Yongliang and Song, Kaitao and Tan, Xu and Li, Dongsheng and Lu, Weiming and Zhuang, Yueting},
  journal={NeurIPS},
  pages={38154--38180},
  year={2023}
}

@inproceedings{
    anthis2025position,
    title={Position: {LLM} Social Simulations Are a Promising Research Method},
    author={Jacy Reese Anthis and Ryan Liu and Sean M Richardson and Austin C. Kozlowski and Bernard Koch and Erik Brynjolfsson and James Evans and Michael S. Bernstein},
    booktitle={ICML (Position Paper Track)},
    year={2025}
}

@article{gao2023s3,
  title={S3: Social-network simulation system with large language model-empowered agents},
  author={Gao, Chen and Lan, Xiaochong and Lu, Zhihong and Mao, Jinzhu and Piao, Jinghua and Wang, Huandong and Jin, Depeng and Li, Yong},
  journal={arXiv preprint arXiv:2307.14984},
  year={2023}
}

@article{ge2024scaling,
  title={Scaling synthetic data creation with 1,000,000,000 personas},
  author={Ge, Tao and Chan, Xin and Wang, Xiaoyang and Yu, Dian and Mi, Haitao and Yu, Dong},
  journal={arXiv preprint arXiv:2406.20094},
  year={2024}
}

@inproceedings{salemi2024lamp,
  title={Lamp: When large language models meet personalization},
  author={Salemi, Alireza and Mysore, Sheshera and Bendersky, Michael and Zamani, Hamed},
  booktitle={ACL},
  pages={7370--7392},
  year={2024}
}

@inproceedings{zhou2024sotopia,
  title = {SOTOPIA: Interactive Evaluation for Social Intelligence in Language Agents},
  author = {Zhou, Xuhui and Zhu, Hao and Mathur, Leena and Zhang, Ruohong and Qi, Zhengyang and Yu, Haofei and Morency, Louis-Philippe and Bisk, Yonatan and Fried, Daniel and Neubig, Graham and Sap, Maarten},
  booktitle = {ICLR},
  year = {2024}
}

@inproceedings{li2024econagent,
  title={Econagent: large language model-empowered agents for simulating macroeconomic activities},
  author={Li, Nian and Gao, Chen and Li, Mingyu and Li, Yong and Liao, Qingmin},
  booktitle={ACL},
  pages={15523--15536},
  year={2024}
}

@article{williams2023epidemic,
  title={Epidemic modeling with generative agents},
  author={Williams, Ross and Hosseinichimeh, Niyousha and Majumdar, Aritra and Ghaffarzadegan, Navid},
  journal={arXiv preprint arXiv:2307.04986},
  year={2023}
}

@article{zhang2024electionsim,
  title={Electionsim: Massive population election simulation powered by large language model driven agents},
  author={Zhang, Xinnong and Lin, Jiayu and Sun, Libo and Qi, Weihong and Yang, Yihang and Chen, Yue and Lyu, Hanjia and Mou, Xinyi and Chen, Siming and Luo, Jiebo and others},
  journal={arXiv preprint arXiv:2410.20746},
  year={2024}
}

@inproceedings{yuan2024evaluating,
  title={Evaluating Character Understanding of Large Language Models via Character Profiling from Fictional Works},
  author={Yuan, Xinfeng and Yuan, Siyu and Cui, Yuhan and Lin, Tianhe and Wang, Xintao and Xu, Rui and Chen, Jiangjie and Yang, Deqing},
  booktitle={EMNLP},
  pages={8015--8036},
  year={2024}
}

@article{wang2025user,
  title={User behavior simulation with large language model-based agents},
  author={Wang, Lei and Zhang, Jingsen and Yang, Hao and Chen, Zhi-Yuan and Tang, Jiakai and Zhang, Zeyu and Chen, Xu and Lin, Yankai and Sun, Hao and Song, Ruihua and others},
  journal={ACM Transactions on Information Systems},
  volume={43},
  number={2},
  pages={1--37},
  year={2025},
  publisher={ACM New York, NY}
}

@article{tu2023characterchat,
  title={Characterchat: Learning towards conversational ai with personalized social support},
  author={Tu, Quan and Chen, Chuanqi and Li, Jinpeng and Li, Yanran and Shang, Shuo and Zhao, Dongyan and Wang, Ran and Yan, Rui},
  journal={arXiv preprint arXiv:2308.10278},
  year={2023}
}

@article{schwitzgebel2023creating,
  title={Creating a Large Language Model of a Philosopher},
  author={Schwitzgebel, Eric and Schwitzgebel, David and Strasser, Anna},
  journal={arXiv preprint arXiv:2302.01339},
  year={2023}
}

@inproceedings{shao2023character,
  author       = {Yunfan Shao and
                  Linyang Li and
                  Junqi Dai and
                  Xipeng Qiu},
  title        = {Character-LLM: {A} Trainable Agent for Role-Playing},
  booktitle    = {{EMNLP}},
  pages        = {13153--13187},
  year         = {2023}
}

@inproceedings{sun2025identity,
  author       = {Libo Sun and
                  Siyuan Wang and
                  Zhongyu Wei},
  title        = {Identity-Driven Hierarchical Role-Playing Agents},
  booktitle    = {{NLPCC}},
  pages        = {403--417},
  year         = {2025}
}

@article{jang2023personalized,
  title={Personalized soups: Personalized large language model alignment via post-hoc parameter merging},
  author={Jang, Joel and Kim, Seungone and Lin, Bill Yuchen and Wang, Yizhong and Hessel, Jack and Zettlemoyer, Luke and Hajishirzi, Hannaneh and Choi, Yejin and Ammanabrolu, Prithviraj},
  journal={arXiv preprint arXiv:2310.11564},
  year={2023}
}

@inproceedings{shea2023building,
  author       = {Ryan Shea and
                  Zhou Yu},
  title        = {Building Persona Consistent Dialogue Agents with Offline Reinforcement Learning},
  booktitle    = {{EMNLP}},
  pages        = {1778--1795},
  year         = {2023}
}

@article{wang2024towards,
  title={Towards objectively benchmarking social intelligence for language agents at action level},
  author={Wang, Chenxu and Dai, Bin and Liu, Huaping and Wang, Baoyuan},
  journal={arXiv preprint arXiv:2404.05337},
  year={2024}
}

@inproceedings{du2023improving,
  title={Improving factuality and reasoning in language models through multiagent debate},
  author={Du, Yilun and Li, Shuang and Torralba, Antonio and Tenenbaum, Joshua B and Mordatch, Igor},
  booktitle={ICML},
  pages={11733--11763},
  year={2023}
}

@article{xu2023exploring,
  title={Exploring large language models for communication games: An empirical study on werewolf},
  author={Xu, Yuzhuang and Wang, Shuo and Li, Peng and Luo, Fuwen and Wang, Xiaolong and Liu, Weidong and Liu, Yang},
  journal={arXiv preprint arXiv:2309.04658},
  year={2023}
}

@article{wang2023avalon,
  title={Avalon's game of thoughts: Battle against deception through recursive contemplation},
  author={Wang, Shenzhi and Liu, Chang and Zheng, Zilong and Qi, Siyuan and Chen, Shuo and Yang, Qisen and Zhao, Andrew and Wang, Chaofei and Song, Shiji and Huang, Gao},
  journal={arXiv preprint arXiv:2310.01320},
  year={2023}
}

@article{li2024agent,
  title={Agent hospital: A simulacrum of hospital with evolvable medical agents},
  author={Li, Junkai and Lai, Yunghwei and Li, Weitao and Ren, Jingyi and Zhang, Meng and Kang, Xinhui and Wang, Siyu and Li, Peng and Zhang, Ya-Qin and Ma, Weizhi and others},
  journal={arXiv preprint arXiv:2405.02957},
  year={2024}
}

@inproceedings{he2024agentscourt,
  title={Agentscourt: Building judicial decision-making agents with court debate simulation and legal knowledge augmentation},
  author={He, Zhitao and Cao, Pengfei and Wang, Chenhao and Jin, Zhuoran and Chen, Yubo and Xu, Jiexin and Li, Huaijun and Liu, Kang and Zhao, Jun},
  booktitle={{EMNLP} (Findings)},
  pages={9399--9416},
  year={2024}
}

@article{koster2022spurious,
  title={Spurious normativity enhances learning of compliance and enforcement behavior in artificial agents},
  author={K{\"o}ster, Raphael and Hadfield-Menell, Dylan and Everett, Richard and Weidinger, Laura and Hadfield, Gillian K and Leibo, Joel Z},
  journal={Proceedings of the National Academy of Sciences},
  volume={119},
  number={3},
  pages={e2106028118},
  year={2022},
  publisher={National Academy of Sciences}
}

@article{schelling1971dynamic,
  title={Dynamic models of segregation},
  author={Schelling, Thomas C},
  journal={Journal of mathematical sociology},
  volume={1},
  number={2},
  pages={143--186},
  year={1971},
  publisher={Taylor \& Francis}
}

@article{xie2024can,
  title={Can large language model agents simulate human trust behavior?},
  author={Xie, Chengxing and Chen, Canyu and Jia, Feiran and Ye, Ziyu and Lai, Shiyang and Shu, Kai and Gu, Jindong and Bibi, Adel and Hu, Ziniu and Jurgens, David and others},
  journal={NeurIPS},
  pages={15674--15729},
  year={2024}
}

@article{mozikov2024good,
  title={The good, the bad, and the hulk-like gpt: Analyzing emotional decisions of large language models in cooperation and bargaining games},
  author={Mozikov, Mikhail and Severin, Nikita and Bodishtianu, Valeria and Glushanina, Maria and Baklashkin, Mikhail and Savchenko, Andrey V and Makarov, Ilya},
  journal={arXiv preprint arXiv:2406.03299},
  year={2024}
}

@article{chaudhary2024large,
  title={Large language models as instruments of power: New regimes of autonomous manipulation and control},
  author={Chaudhary, Yaqub and Penn, Jonnie},
  journal={arXiv preprint arXiv:2405.03813},
  year={2024}
}

@inproceedings{chuang2024wisdom,
  title={The Wisdom of Partisan Crowds: Comparing Collective Intelligence in Humans and LLM-based Agents},
  author={Chuang, Yun-Shiuan and Harlalka, Nikunj and Suresh, Siddharth and Goyal, Agam and Hawkins, Robert and Yang, Sijia and Shah, Dhavan and Hu, Junjie and Rogers, Timothy T},
  booktitle={CogSci},
  volume={46},
  year={2024}
}

@inproceedings{wang2024rolellm,
  author       = {Noah Wang and
                  Zhongyuan Peng and
                  Haoran Que and
                  Jiaheng Liu and
                  Wangchunshu Zhou and
                  Yuhan Wu and
                  Hongcheng Guo and
                  Ruitong Gan and
                  Zehao Ni and
                  Jian Yang and
                  Man Zhang and
                  Zhaoxiang Zhang and
                  Wanli Ouyang and
                  Ke Xu and
                  Wenhao Huang and
                  Jie Fu and
                  Junran Peng},
  title        = {RoleLLM: Benchmarking, Eliciting, and Enhancing Role-Playing Abilities of Large Language Models},
  booktitle    = {{ACL} (Findings)},
  pages        = {14743--14777},
  year         = {2024}
}

@article{park2024diminished,
  title={Diminished diversity-of-thought in a standard large language model},
  author={Park, Peter S and Schoenegger, Philipp and Zhu, Chongyang},
  journal={Behavior Research Methods},
  volume={56},
  number={6},
  pages={5754--5770},
  year={2024},
  publisher={Springer}
}

@article{ju2024sense,
  title={Sense and sensitivity: evaluating the simulation of social dynamics via large language models},
  author={Ju, Da and Williams, Adina and Karrer, Brian and Nickel, Maximilian},
  journal={arXiv preprint arXiv:2412.05093},
  year={2024}
}

@inproceedings{gao2023livechat,
  author       = {Jingsheng Gao and
                  Yixin Lian and
                  Ziyi Zhou and
                  Yuzhuo Fu and
                  Baoyuan Wang},
  title        = {LiveChat: {A} Large-Scale Personalized Dialogue Dataset Automatically Constructed from Live Streaming},
  booktitle    = {{ACL}},
  pages        = {15387--15405},
  year         = {2023}
}

@inproceedings{chen2023large,
  title={Large language models meet harry potter: A dataset for aligning dialogue agents with characters},
  author={Chen, Nuo and Wang, Yan and Jiang, Haiyun and Cai, Deng and Li, Yuhan and Chen, Ziyang and Wang, Longyue and Li, Jia},
  booktitle={{EMNLP} (Findings)},
  pages={8506--8520},
  year={2023}
}

@inproceedings{liu2022improving,
  title={Improving personality consistency in conversation by persona extending},
  author={Liu, Yifan and Wei, Wei and Liu, Jiayi and Mao, Xianling and Fang, Rui and Chen, Dangyang},
  booktitle={CIKM},
  pages={1350--1359},
  year={2022}
}

@inproceedings{li2024steerability,
  title={The steerability of large language models toward data-driven personas},
  author={Li, Junyi and Peris, Charith and Mehrabi, Ninareh and Goyal, Palash and Chang, Kai-Wei and Galstyan, Aram and Zemel, Richard and Gupta, Rahul},
  booktitle={NAACL},
  pages={7290--7305},
  year={2024}
}

@inproceedings{ahn2023mpchat,
  title={MPCHAT: Towards Multimodal Persona-Grounded Conversation},
  author={Ahn, Jaewoo and Song, Yeda and Yun, Sangdoo and Kim, Gunhee},
  booktitle={ACL},
  pages={3354--3377},
  year={2023}
}

@inproceedings{chawla2023selfish,
  title={Be selfish, but wisely: Investigating the impact of agent personality in mixed-motive human-agent interactions},
  author={Chawla, Kushal and Wu, Ian and Rong, Yu and Lucas, Gale and Gratch, Jonathan},
  booktitle={EMNLP},
  pages={13078--13092},
  year={2023}
}

@article{light2023avalonbench,
  title={Avalonbench: Evaluating llms playing the game of avalon},
  author={Light, Jonathan and Cai, Min and Shen, Sheng and Hu, Ziniu},
  journal={arXiv preprint arXiv:2310.05036},
  year={2023}
}

@inproceedings{
li2025simulating,
title={Simulating Society Requires Simulating Thought},
author={Chance Jiajie Li and Jiayi Wu and Zhenze Mo and Ao Qu and Yuhan Tang and Kaiya Ivy Zhao and Yulu Gan and Jie Fan and Jiangbo Yu and Jinhua Zhao and Paul Pu Liang and Luis Alberto Alonso Pastor and Kent Larson},
booktitle={{NeurIPS} (Position Paper Track)},
year={2025}
}






\newpage
\appendix
\section{More Details of Our Scope}
\label{sec:appendix}
Social simulation generally denotes the use of computer-based simulations to model social processes. Its core objective is to computationally emulate human individuals’ cognition, decision-making, and social interactions in an artificial yet computationally tractable environment, so as to generate, explain, or predict emergent social phenomena at the collective level \cite{gilbert2005simulation}. Beyond scalability and cost considerations, social simulation plays a fundamental methodological role in social science by enabling controlled, repeatable, and counterfactual analysis of complex social systems. Traditional empirical approaches in social science, such as surveys  \cite{devaus2013surveys}, laboratory experiments  \cite{falk2009lab}, and field studies  \cite{baldassarri2017field}, primarily observe social phenomena as they unfold, but are often constrained in their ability to systematically manipulate conditions, explore rare or high-risk scenarios, or disentangle causal mechanisms underlying collective outcomes  \cite{bruch2015agent}. In contrast, social simulation allows researchers to explicitly specify micro-level behavioral assumptions and interaction rules, and to examine how alternative individual decisions, institutional designs, or environmental conditions give rise to divergent macro-level dynamics. This capability is particularly valuable for studying policy interventions \cite{faucher2022agent}, public opinion dissemination \cite{li2020opinion}, social norms \cite{chen2012agent}, and strategic behavior \cite{granha2022opinion}, where real-world experimentation may be infeasible, unethical, or prohibitively costly. By complementing empirical methods with a flexible and interpretable computational testbed, social simulation provides social science with a powerful tool for theory exploration, hypothesis generation, and ex ante policy assessment.

In recent years, this methodological landscape has driven growing interest in using large language model (LLM)-integrated AI agents for social simulation \cite{gao2024large,mou2024individual,piao2025agentsociety}. Traditionally, agent-based social simulation has relied on hand-crafted rules  \cite{wolfram1984cellular}, utility functions  \cite{vinkovic2006physical}, or simplified behavioral equations \cite{helbing1995social} to model individual decision-making and interaction. While such approaches offer transparency and analytical control, they often impose strong structural assumptions on human behavior, leading to overly stylized dynamics that struggle to capture the richness, heterogeneity, and context sensitivity of real human cognition. Moreover, rule-based agents typically exhibit limited adaptability, requiring extensive manual redesign when transferred across scenarios, populations, or intervention settings \cite{gao2024large}. The emergence of LLMs has introduced a paradigm shift by enabling agents with substantially improved natural language understanding, generation, and reasoning capabilities \cite{achiam2023gpt,touvron2023llama,guo2025deepseek}. It is widely argued that such agents have the potential to engage in open-ended communication, negotiation, persuasion, and coordination through natural language, and to flexibly adopt diverse roles or personas when conditioned on contextual prompts \cite{park2023generative,shinn2023reflexion}. Under this prevailing assumption, AI agents are often viewed as providing a scalable and expressive substrate for modeling complex social interactions across a wide range of scenarios, making them an appealing foundation for contemporary social simulation in domains such as policy analysis, behavioral forecasting, and opinion dynamics.

In this paper, we focus on the use of (LLM-integrated) AI agents \cite{gao2024large,mou2024individual,anthis2025position} for social simulation, specifically in settings where agents are employed as proxies for human individuals and simulation outcomes are interpreted at the collective level. Our analysis centers on the methodological assumptions and implications of adopting such agents to model social behavior and interaction, rather than mainly on task performance or predictive accuracy. We do not aim to provide a comprehensive survey of social simulation, nor do we consider non–agent-based approaches \cite{johnson2019hidden} beyond brief methodological comparison. In addition, we do not study the intrinsic ``personality'' of LLMs themselves \cite{jiang2023evaluating,bhandari2025evaluating,serapio2025psychometric} or the interpretability of their behaviors \cite{shanahan2023role,zou2023representation,sharma2024towards}, nor do we examine applications of AI agents beyond social simulation, such as agents designed for scientific research \cite{boiko2023autonomous,romera2024mathematical,m2024augmenting}, software development \cite{hong2023metagpt,yang2024swe,qian2024chatdev}, or other task-oriented workflows \cite{yao2023react,shen2023hugginggpt,wang2023voyager}. This scoped focus allows us to critically examine prevailing assumptions and overlooked issues of AI agents specifically within the context of social simulation.





\end{document}